# Asymmetric Aging Effect on Modern Microprocessors

Freddy Gabbay and Avi Mendelson, *Fellow, IEEE*

*Abstract*—**Reliability, a crucial requirement in any modern microprocessor, assures correct execution over its lifetime. As mission critical components are becoming common in commodity systems, e.g., control of autonomous cars, the demand for reliable processing continues to grow. The latest process technologies have aggravated the situation by causing microprocessors to be highly vulnerable to reliability concerns. This paper examines the asymmetric aging phenomenon, which is a major reliability concern in advanced process nodes. In this phenomenon, logical elements and memory cells suffer from unequal timing degradation over lifetime and, consequently, introduce reliability concerns. Thus far, most studies approached asymmetric aging from a circuit or physical design viewpoint, but these solutions were quite limited and suboptimal. In this paper, we introduce an asymmetric aging-aware microarchitecture that aims to reduce the phenomenon's impact. The study is mainly focused on the following subsystems: execution units, register files and memory hierarchy. Our experiments indicate that the proposed solutions incur minimal overhead while significantly mitigating asymmetric aging stress.**

*Index Terms*—**Asymmetric Aging, Reliability, Bias Temperature Instability, Reliability-Aware Architecture**

## I. INTRODUCTION

CHIP reliability, a crucial requirement for modern microprocessors, is essential to assure correct execution over a system's lifetime. New mission-critical computation-intensive applications (e.g., autonomous vehicles, data centers, cloud computing, life-support systems, etc.) impose strict requirement demands on reliability through a product's lifetime and operating conditions. For example, the automotive industry expects an integrated circuits (IC) to function reliably for 10–15 years at a given temperature (usually about 125°C [1,2] and under various workloads. In data centers, the requirements are slightly relaxed but remain challenging: the lifetime requirement stipulates at least ten years, but under conditions where the temperature can range from 105 to 110°C with arbitrary workloads. None of these reliability-sensitive applications can afford microprocessor faults caused by reliability issues. While reliability requirements have been substantially heightened, advanced FinFET VLSI technologies (28nm and lower) have become highly suspectable to reliability issues and, in particular, to transistor aging. Transistor aging is the deterioration process of transistors due to charge carriers trapped at the dielectric insulator of a transistor gate. This phenomenon is induced by Hot Carrier Injection (HCI) and Bias Temperature Instability (BTI) effects that are further discussed in Section 2. The common approach to handling such degradation in digital circuits is to provide extra timing margins to the clock-cycle time, hence, taking the timing degradation into account (as a result, microprocessors incur performance degradation). One may consider this to be an adequate solution as long as the aging is symmetric, i.e., all transistors age at the same rate. Unfortunately, many digital circuits may incur asymmetric transistor aging. Consequently, different paths in circuits may incur uneven degradation that may result in critical timing violations. Asymmetric aging is mainly induced by the BTI effect ([3–5]) as a result of applying constant voltage to transistor gates for long periods. The period of time required for the transistor to incur such degradation may vary between tens of seconds up to several weeks ([4]). Accordingly, even if the asymmetric delay shift is relatively small, critical timing constraints in the logical circuit may not be met. This observation makes logical elements that are maintained under constant logical values for long periods susceptible to severe reliability issues. The problem is even more severe for logical structures such as register files and memory arrays where a single element failure may fail the whole module and, as a result, data integrity may become unreliable. Past studies indicated that from the architectural point of view, asymmetric aging is commonly induced by dynamic-power saving techniques, e.g., clock gating, which enforce a static state on logical circuits and, consequently, they incur BTI stress ([6]).

Many of the prior studies, described in Section 2.3, approached asymmetric aging from the physical design point of view. Such approaches were not straightforward since they involved highly complex simulations and analysis methods. Even many of the common electronic design automation (EDA) tools today cannot yet handle such tasks and, in particular, are very limited in their ability to analyze very large-scale circuits that could employ billion of transistors.

In this paper, our main focus is minimizing microprocessor susceptibility to asymmetric aging induced by static BTI stress applied for long periods. We present three novel mechanisms to mitigate asymmetric aging under constant stress conditions. The first mechanism avoids asymmetric aging in execution units that are under BTI stress through periodic injection of pseudorandom data at low rates. The second mechanism deals with asymmetric aging avoidance in architectural and control registers by employing periodic shifts and remapping of register identifiers. The third mechanism handles asymmetric aging in memory systems using a new approach that combines swap-shift of cache sets and pseudorandom data generation.

The remainder of this paper is organized as follows: Section 2 introduces asymmetric aging reliability challenges and reviews previous works. Section 3 describes our asymmetric aging observations and Section 4 presents our proposed microarchitecture enhancements for modern microprocessors accompanied by experimental results. Finally, Section 5 summarizes the study and suggests directions for future research work.

## II. ASYMMETRIC AGING

The susceptibility of modern process technologies to reliability-related issues has grown dramatically. Starting at 28nm process technology and below (16, 7, 5, and 3nm), design efforts dedicated to reliability have substantially increased. The design community has mainly tried to enhance the synthesis and place-and-route flows to minimize and eliminate reliability-related issues. Such flows involve substantial design efforts and, in many cases, require multiple iterations to make the IC comply with the design rules (also known as the "sign-off process").

We now describe the asymmetric aging effect and thereafter provide an overview of previous studies.

### A. Asymmetric Transistor Aging

Transistor aging is the deterioration process of transistors, residing in logical gates and memory elements ([8, 9]), due to charge carriers from the transistor inversion channel being trapped at the dielectric insulator of a transistor gate. There are two physical mechanisms that cause charge carriers to be trapped: 1. Hot Carrier Injection (HCI), which involves charge carriers that flow from the transistor source to the drain; the charge carriers may get trapped in the gate oxide due to excessive energy levels. 2. Bias Temperature Instability (BTI), where charge carriers are also caught in the dielectric gate insulator, but this time no current flow is required between the source and drain of the transistor; rather, they may get caught whenever voltage is applied to the transistor gate. When gate voltage is removed after a short period of time (<10s), the damage is partially reversible, and part of the charge carriers is detached.

Both BTI and HCI increase the transistor threshold voltage, reduce charge carriers' mobility in the channel, and mandate a higher voltage to switch on the transistor. In addition, they also slow down transistor speed due to the degradation in the transistor current. As a result, ICs may experience major frequency degradation over their lifetime. Various methods for dealing with aging affects have been offered and these are physical design or circuit-based solutions ([9–11]). The most common approach is to provide extra margins for the clock-cycle time to compensate for the lifetime performance degradation.

Recent studies have discovered that the degradation due to aging may not be uniformly distributed. This may happen in the following scenarios:

1) Inside a logical cell when p-devices and n-devices age unequally and, as a result, rising and falling transient edges may experience different delay shifts.
2) Between different paths in a logical circuit [7, 12], which incur uneven aging degradation, may result in critical timing constraint violations.

When such violations involve setup timing constraints, they can be mitigated by reducing the clock frequency; however, when hold constraints are violated, the circuit will incur severe reliability issues that cannot be mitigated. This phenomenon, referred to as "asymmetric aging", has become a major reliability concern in mission critical systems.

Asymmetric aging is induced as a result of static stress applied to logic gates or memories for long periods, which may vary between tens of seconds up to several weeks ([7]). BTI has been found to be the main contributor to this phenomenon that may affect both p-type (known as NBTI) and n-type (PBTI) transistors. NBTI exhibits higher impact by several orders of magnitudes relative to PBTI, though in advanced process technologies PBTI was shown to also have considerable impact. As a result of the BTI effect, logical paths that are under different static stress will age asymmetrically and introduce new timing violations that cannot be identified by conventional timing verification methods.

Asymmetric aging is highly complex to model, analyze, predict, and avoid in very large-scale ICs and, therefore, it has become a major reliability issue. In addition, timing analyses that takes into account the asymmetric aging effect are non-trivial as they depend not only on the mode of operation (static vs. dynamic stress) but also on the operating conditions and technology specifications. Conventional timing verification tools ([13]) lack any information related to the lifetime activation modes of the digital circuit, e.g., standby modes, constant values and activation of clock gates that are applied for long periods. The complexity of handling asymmetric aging also exists in reliability qualification tests such as High-Temperature Operating Life (HTOL) ([14]), which are commonly used by the industry to test the lifetime reliability of ICs. HTOL tests accelerate the aging degradation by running circuits at high temperature and under high voltage; however, they assume symmetric aging and use designated circuitry to assure that all logical paths are kept toggling through the test period in order to avoid BTI stress. Such limitations, which are major concerns in assuring IC reliability, motivate us to explore new approaches to mitigate the asymmetric aging effect.

From an architectural point of view, asymmetric aging in many cases is a result of dynamic power saving techniques that enforce a static state on logical circuits, so that they incur a BTI effect. This effect is demonstrated, in Fig. 1, on five typical scenarios common to modern microprocessors and SoCs. Fig. 1 (a) depicts a bit cell circuit that is the basic circuit of SRAMs. A bit cell consists of two cross-coupled inverters and pass transistors that provide access to the bit cell. No matter what data is stored in the bit cell, there is always an inverter with a logical 1 in its input while the other inverter has 0. Whenever a 0 is applied to a logical gate, the p-type transistor will suffer from NBTI stress and, as a result, the inverter will age asymmetrically in respect to the other inverter.

Fig. *1* (b) illustrates a similar phenomenon in an SR latch. SR latches are the building blocks of master-slave D flip-flops registers that are broadly used in digital circuits. Whenever the latch maintains the same storage value (SR=00), it similarly behaves like the bit-cell cross-coupled inverters, which consequently incur asymmetric aging.

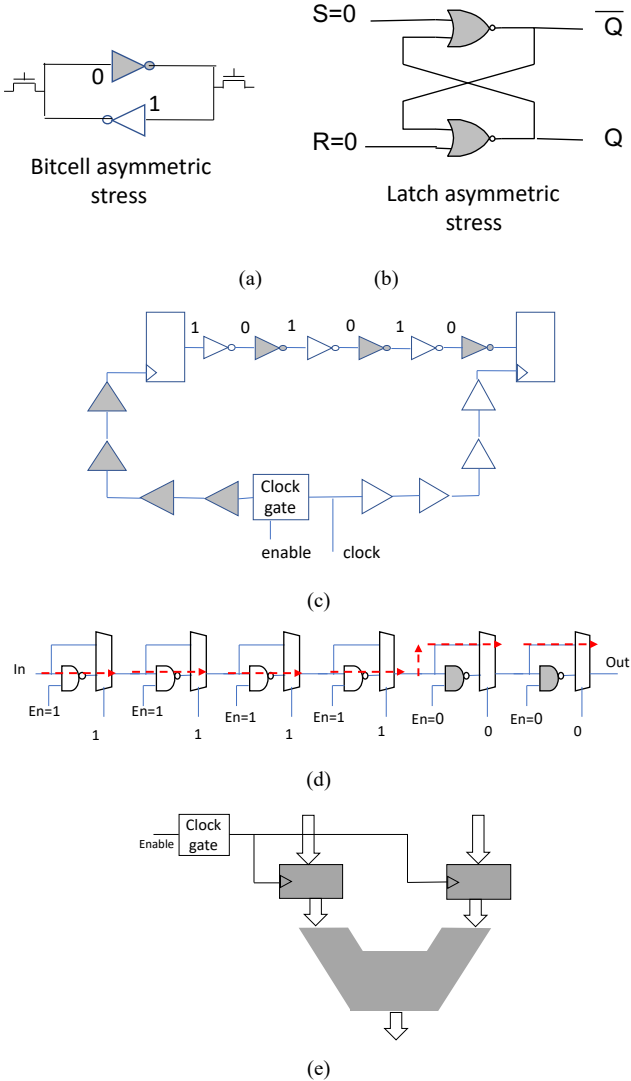

(a)          (b)

(c)

(d)

(e)

Fig. 1. Asymmetric aging in (a) a bit cell, (b) an SR latch, (c) launch and capture logical paths, (d) a delay chain and (e) an execution unit.

Another example of a general synchronous digital circuit is shown in Fig. *1* (c) where the launch clock path is controlled by a clock gate while the capture clock path is free running. Such a circuit implementation is common in microprocessors to save dynamic power, particularly when the launch path changes infrequently. If the clock gate disables the launch clock for a long time, static stress will be applied to all the clock buffers in the launch clock path. Hence, they will experience a higher delay shift relative to the capture path, resulting in an unbalanced clock tree between the launch and captured clock endpoints. Such a design will incur a degradation in setup slack that may lead to setup violation due to the late arrival of the launch clock edge. If the clock gate is placed in the capture path, it may suffer from hold violation. The logical data path between the flops, illustrated in Fig. 1 (c) by a chain of inverters, may suffer from timing violations as well. Whenever the launch flop output value is kept static for a long time, the static stress will emerge through all the inverters in the chain. In particular, all inverters with 0 on their gate will stress their p-type transistors and, as a result, will introduce a higher raise delay shift in respect to the other inverters.

The example illustrated by Fig. *1* (d) is a delay chain circuit that is commonly used in I/O interfaces such as DDR DRAM to adjust timing delays. Such a circuit consists of a controllable chain of delay gates where the programable delay is determined either by production testing or at boot time based on the training sequence. The unused delay gates will be excluded from the delay path and will remain static to save dynamic power. As a result, these delay gates will incur asymmetric aging and if they are used in the future, they will exhibit severe reliability issues due to the delay shift between the rise and fall times. In certain cases, the signal will not even be able to propagate through the chain when the delay shift becomes high.

The last example depicted by Fig. *1* (e) shows an idle data path of an execution unit (e.g., an FP adder or multiplier) that is under static stress. The inputs to the block are stored by clock-gated registers and when the unit is idle, the clock is disabled.

### B. Prior Studies

Many of the prior works approached asymmetric aging from the physical design point of view. Such an approach is not straightforward, since the process of simulating, analyzing and fixing asymmetric aging issues in large-scale circuits is highly complex ([15]), with only few EDA tools (e.g., BERT, RelXpert) today having such a limited capability ([16, 17]). The strategies that were suggested attempted to cope with the problem from various directions. One approach was to enhance the process node to reduce the impact of the BTI effect. This, however, became highly challenging due to the down-scaling dimension of the gate oxide. Traditional approaches relied on taking margins in timing closures for both setup and hold that would take into account the asymmetric aging effect. This was found to necessitate a highly complex analysis and, in many cases, ended up in overdesign. Other studies attempted to model and predict the degradation as a result of NBTI ([3–5, 7, 12, 13, 18–20]) and suggested various solutions such as transistor sizing, $V_{DD}$ tuning, duty cycle reduction and also decreasing the transistor channel length. Agrawal et al. in [21] presented a mechanism for circuit failure prediction by collecting data from special sensors placed in different locations in the silicon die. Their results indicate that by using these sensors, they can reduce the conservative margins used by the traditional design flows and improve chip performance. Further studies ([22–25]) also introduced methods for analyzing digital circuits and identifying critical gates that are the most susceptible to NBTI stress. This was done by employing an aging model (consisting of BTI-aware libraries) and an aging-aware timing analysis.

Several other studies ([26–29]) suggested applying power gating and drowsy memory to mitigate BTI stress in memories. Power gating employs transistor switches that powers down the entire SRAM, and as a result, eliminates the BTI stress all through the shutdown process. Power gating, however, has several limitations. First, when applied, the entire content of the SRAM is lost and cannot be recovered when the SRAM is switched back on. In addition, power gating introduces a performance overhead when the memory is switched back on, and, therefore, it is inefficient when the memories become idle for a long time. Drowsy memory is another technique for power saving in SRAMs. When the memory is entered into low power mode ("drowsy" mode), the memory voltage is reduced to the minimum retention voltage that allows bit cells to continue maintaining their values safely without being accessible. When the memory is accessed, the voltage is switched back to nominal voltage. Throughout the drowsy state, the BTI stress on bit cells is reduced since $V_{DD}$ reduction also reduces the gate voltage.

Other studies examined asymmetric aging on different logical circuits that can suffer from significant NBTI degradation. Velamala et al. ([7, 18]) introduced the NBTI effect on DDR delay chains and SRAM internal circuits. Another SRAM asymmetric aging study by [30] suggested an on-chip reliability monitor to measure SRAM BTI impact. Their study found that SRAMs can incur significant internal clock duty-cycle shifts in read operations as a result of asymmetric aging. Yan et al. ([31]) studied NBTI impact on master-slave D flip-flops under different duty-cycle assumptions.

While many studies attempted to cope with asymmetric aging from the physical design point of view, only a limited number of works examined this phenomenon from an architectural point of view. Firouzi et al. suggested a NOP instruction insertion to reduce the impact of NBTI on the execution stage of MIPS processors ([32]). They found that such NBTI degradation is dominated by the values of source operands rather than opcodes and suggested software and hardware approaches to relax NBTI stress by using different variations of NOP instructions. This method was found to provide limited improvement in processors that employ a negligible number of NOP instructions ([33]). In addition, it may increase GPR register utilization and, as a result, reduce the number of available registers for the application. Abbas et al. suggested running anti-aging programs instead of idle tasks when the processor is not utilized ([33]). The anti-aging programs generate specific value patterns to repair the BTI asymmetric aging effect in the execution unit combinatorial circuits. This technique was efficient; however, it required complex analysis of the critical paths and the requisite anti-aging values. Moreover, it was limited to handling the execution stage combinatorial circuits only and assumed a scalar processor. Note that in the case of out-of-order processors, such techniques may become limited in mapping the anti-aging patterns to the multiple execution units. Chen et al. examined the performance degradation due to asymmetric aging in multicore systems where processors may be asymmetrically aged due to different workloads and utilizations ([34]). They suggested reserving certain cores at early stages of the system lifetime to be used for executing critical missions at late stages.

Field Programable Gate Array (FPGA) devices may also be highly susceptible to NBTI. Unused FPGA logic can suffer from long constant logical stress and, when such logic is used again, it may incur asymmetric aging degradation. A technique to reduce the impact of asymmetric aging on FPGA was introduced by [35] who suggested bundling unused FPGA elements in logical chains and toggling them at low rates to prevent the constant NBTI stress.

Other studies proposed solutions for asymmetric aging in the processor's memory system. As previously discussed, SRAM memories are highly susceptible to asymmetric aging due to the basic structures of bit cells, and when a bit cell stores static values for a long period, e.g., startup value (SUV), secret keys etc., severe reliability concerns induced by asymmetric aging may arise ([36]). Various techniques aimed at mitigating SRAM asymmetric aging degradation by balancing the signal probability of 0 and 1 states. Kumar et al. suggested a periodic bit-flipping process for the first level cache where every cache line is periodically read, inverted and written back to the cache ([37]). When managed by software, this mechanism introduced significant performance overhead due to the disruption of real cache accesses by the processor. A hardware-based approach that maintains the bit flipping locally in the cache SRAM was reported to reduce the memory access overhead. Gebregiorgis et al. proposed a self-controlled bit-flipping (SCF) method that performs the bit flipping upon every I-cache and D-cache line write access and avoids processor interruptions ([38]). This approach was found to be limited when the cache miss rate was low or when the rate of writes small. Duan et al. introduced a cell flipping technique with distributed refresh phases (CFDR) to reduce the NBTI effect in the I-cache ([39]). This technique flips and refreshes I-cache blocks at a certain refresh rate by scanning cache blocks from a lower to higher index. Since the refresh rate is relatively very low, the processor disruption time becomes negligible. Such a method was reported to introduce a 125% lifetime improvement.

Several past studies attempted to handle cache asymmetric aging using a different approach. Calimera et al. ([40]) suggested probing and scrambling functions for cache re-indexing in fine-grain and coarse-grain modes of operations. Whenever the indexing function was changed, it required flushing the cache content. Their study was examined under two different cache power saving modes: power gating and drowsy mode. The results showed that their method, when used with power saving techniques, can extend cache lifetime up to six-fold. Nevertheless, it may incur a timing overhead on the cache access timing path.

## III. Asymmetric Aging Enhancements in Modern Microprocessors

In this section we present new microarchitecture enhancements for modern microprocessors to cope with the asymmetric aging problem. We start this section by presenting our simulation environment and then we present our experimental observation of asymmetric aging in existing microprocessor architectures. Finally, we suggest microarchitecture enhancements to mitigate asymmetric aging and present our experimental observations of them.

### A. Experimental Environment

For this study, we used the Sniper x86-64 microarchitecture simulator [41]. We modified the simulation platform and added the needed mechanisms to model the behavior and measure the characteristics required for our experiments. The simulation environment included both a detailed cycle-level x86 core model and a memory system. TABLE 1 summarizes the configuration of the simulation environment (based on the Intel Gainestown core [42]).

TABLE 1
BASELINE SIMULATION MODEL CONFIGURATION

| Core Model | |
|---|---|
| Frequency | 2.66 GHz |
| Execution units [time] | 3 ALUs [1 cycle], 1 FP add / sub [3 cycles], 1 FP mul /div [5/6 cycles], 1 branch [1 cycle], 1 load unit [1 cycle], 1 store unit [1 cycle] |
| Pipeline | Dispatch width: 4, Out-of-order (instruction window: 128). |
| Memory System Model | |
| Block size | 64 bytes |
| L1-D Cache | 32KB, 8-way, LRU, 4 clock-cycle access time and a throughput period of one cycle. |
| L1-I Cache | 32KB, 4-way, LRU, 4 clock-cycle access time with instruction prefetching and an instruction queue of 16 bytes per cycle throughput |
| L2 Cache | 256KB, 8-way, LRU, 8 clock-cycle access time |
| L3 Cache | 8MB, 16-way, LRU, 30 clock-cycle access time |
| D-TLB | 64 entries, 4-way |
| I-TLB | 128 entries, 4-way |
| S-TLB | 512 entries, 4-way (secondary TLB) |

We used the Spec2017 benchmarks [43, 44] as our workload benchmarks with *ref* inputs and EMBC Coremark (two different runs). Every Spec2017 benchmark was run as a single-core workload in two different regions of interest: during the initialization phase and during the main execution phase (denoted "Init" and "Main", respectively). Each experiment used 10 billion instructions (for both initialization and main execution phases). The EMBC Coremark benchmark was run from the beginning to completion.

### B. Asymmetric Aging Experimental Observations

Our experimental observations focus on three domains in modern microprocessors: CPU architectural registers, execution units, and memory hierarchy and page tables. Our main focus is identifying elements that are under static stress for long duration. We start our examination by inspecting all architectural registers: general-purpose registers, floating point (FP) registers, vector registers and various control registers. Our experimental results are presented in TABLE 2, which summarizes all register groups with static BTI stress, i.e., they are written only once through simulations or not written at all. As can observed, control registers, e.g., CRs and MSRs, incur major BTI stress as they are kept constant through very long execution periods. Additional memory protection, segment registers and debug registers also incur similar BTI stress. Surprisingly, we also observe that even computation-oriented registers such as FP/vector registers as well as temporary registers suffer from constant stress. These observations are highly concerning as they indicate that critical functions, e.g., page table pointers, cache configuration registers, memory protection mechanisms and pure computational values stored in such registers, may experience severe asymmetric aging.

TABLE 2
X86 REGISTERS WITH STATIC STRESS

| Register Group | Registers with constant stress | |
|---|---|---|
| | Spec2017 | Coremark |
| Memory | BNDCFGU, BNDSTATUS, BND0, BND1, BND2, BND3 | |
| Control | CR0-15, XCR0, TR, TSC, TSCAUX, MXCSR[+] | |
| Debug | DR0-7, ERROR, MSRS | |
| Stack | SSP, IA32_U_CET, STACKPOP | |
| OpMask | K0-7 | |
| X87 | X87CONTROL[+], X87STATUS[+], X87TAG, X87PUSH, X87POP, X87POP2, X87OPCODE, X87LASTCS, X87LASTIP, X87LASTDS, X87LASTDP | |
| Segment | ES, CS, SS, DS, FS, GS, GDTR, LDTR, IDTR, FSBASE, GSBASE | |
| Temporary | TM0-15 | |
| FP Stack | ST0-7 | |
| FP Registers | ZMM10-31 | ZMM16-31, ZMM0-15[++] |

[+]Registers whose values change very seldom.
[++]Registers are static in integer operations.

Our next experiments examine execution units that are under static BTI stress. TABLE 3, which presents our basic observations, shows that FP adder/subtractor and multiplier/divider execution units may incur very long periods of static stress as they are not utilized during the execution in the CoreMark benchmarks and some of the Spec2017 benchmarks that do not use any FP operation (e.g., 602.gcc, 605.mcf, 631.deepsjeng and others). These observations are also highly troubling as they may augur major reliability issues.

TABLE 3
EXECUTION UNITS UNDER STATIC STRESS

Our final observations pertain to the memory hierarchy and the page translation tables. The results are presented in Fig. 2, which illustrates the number of entries with constant stress throughout the entire simulation experiment (either written once or not written at all). Our observations indicate that the L1-I cache is much more susceptible to asymmetric aging, mainly due to a small miss ratio that encourages the line to remain static in the cache. In the L2 cache, we also observe a significant number of lines with static stress, especially in the Coremark benchmarks that have a smaller footprint in respect to the Spec2017 benchmarks (which also introduce a small number of entries with constant stress). The L3 cache is the module with the biggest number of cache lines under constant stress and this is due to either cache entries with low temporal locality that wait for a very long time until they are evicted, or because of unutilized cache lines. Page translation tables also exhibit the existence of a significant number of entries with constant stress.

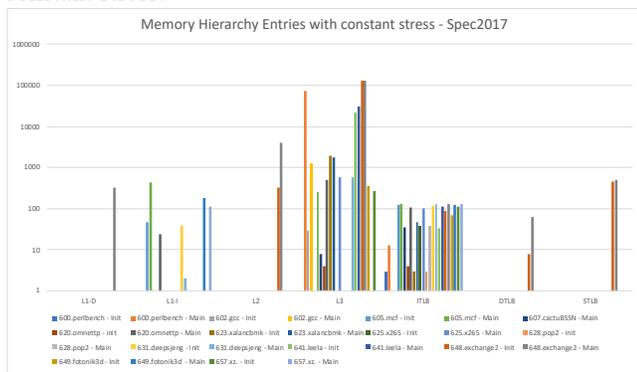

(a)

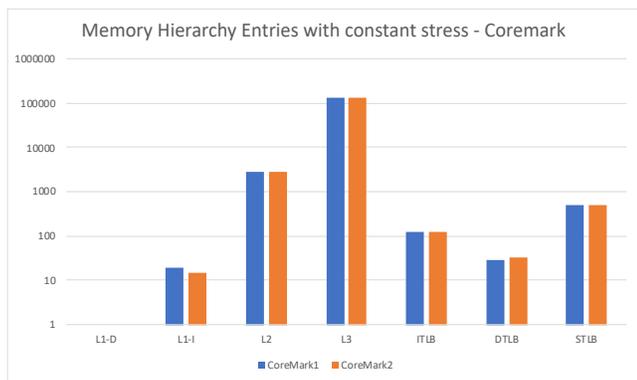

(b)

Fig. 2. Entries in memory hierarchies with static stress: (a) Spec2017 benchmarks and (b) Coremark benchmarks.

It can be observed that the ITLB is very dominant for the Spec2017 and this is due to the high locality of code footprints that consist of many pages that fit into the main memory and, as a result, maintain a significant number of page entries with constant values. In the Coremark benchmark we also observe that both DTLB and STLB maintain a significant number of entries with constant stress. These results raise more concern for the page translation tables residing in main memory that are kept under constant values for long periods. The existence of memory elements under such major constant BTI is very disturbing as bit cells are highly susceptible to BTI stress and any such singular failure may induce a major reliability issue for the entire system. We realize that the likelihood of memory BTI stress in systems that utilize frequent context switches is very low. Nevertheless, machines that run constant jobs, e.g., microcontrollers in embedded systems or database servers, may likely experience the reported observations.

When examining the root cause of these observations from an architectural point of view, we identify three main causes. First, backward compatible features that are not in use by new applications, e.g., x87 control registers, may lead to underutilization and BTI stress. Second, new forward compatible features, e.g., x86 temporary registers, when not efficiently utilized due to legacy considerations, may also generate asymmetric aging concerns. Third, microprocessors are by nature general purpose machines and, consequently, it is highly challenging to inhibit static stress on all their functional units, e.g., FP execution units. The latter observation is also supported by [45] who estimates that 21–50% of microprocessors transistors are underutilized.

## IV. Asymmetric Aging-Aware Microarchitecture Enhancements

This section introduces microarchitectural solutions to reduce the asymmetric aging effect on different microprocess architectural subsystems: execution units, register files, and cache memories.

### A. Execution Units

Based on our experimental observations summarized in Table 3, FP execution units may be under statics stress for long periods. One may argue that in general purpose environments, the use of FP resources may be distributed more evenly, yet in many other environments, such as embedded systems, autonomous cars or even in-memory data servers that run the same workload for very long times, the impact of asymmetric use of the FP execution units may be extremely severe. In this section, we propose a novel scheme to mitigate BTI stress over execution units. Note that although the technique is suggested for FP execution units, it is applicable to any other data path computational module that may be susceptible to asymmetric aging as well. Our scheme utilizes a pseudorandom sequence bit (PRBS) generator that is activated by a slow frequency clock. The PRBS generator generates pseudorandom patterns that are fed into the execution units to prevent extended periods of constant stress. The clock frequency of such a PRBS generator can be in the order of MHz or even lower, to minimize dynamic power overhead. Varieties of PRBSs generators are used for communication and security applications. We examined a simple PRBS circuit ([46]) as part of our study, which introduces very small logic and power overhead while being able to generate random patterns that are sufficient to toggle the execution unit at a low rate.

We examined the effectiveness of the proposed technique on a double precision FP adder design from OpenCores[1]. The block was connected to a PRBS generator as illustrated in Fig.

*3*, synthesized, and the signal probability (the ratio of signal node at the gate level being active) was measured during a simulation of one million clock cycles.

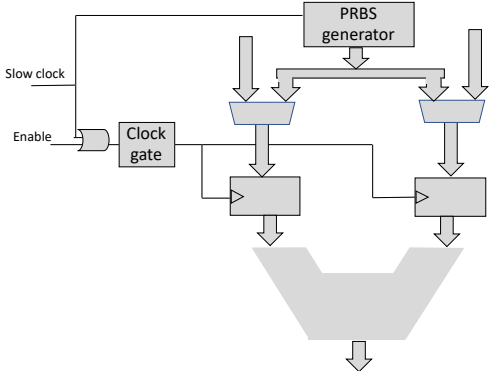

Fig. 3 Asymmetric aging aware scheme for execution units.

The histogram presented in Fig. *4* depicts the signal probability distribution across all the signals in the design. It can be clearly observed that the majority of the signal has a signal probability of approximately 50% due to the values injected by the PRBS circuit. The histogram shows that only a small group of signals could not be toggled effectively and remained static through most of the simulation. We identified that this group is related to the FP addition result shift that involves zero-padding. This can be easily fixed by forcing the PRBS patterns to be injected into this group of signals that, as a result, avoids the constant state.

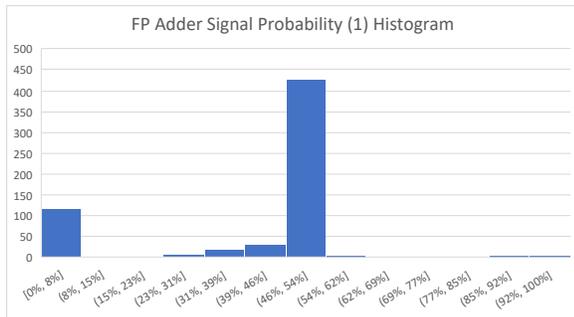

Fig. 4. FP adder toggle rate histogram.

As part of our examination, we also summarized (Table 4) the overall overhead on 28nm-process technology using high VT (HVT) standard cells, which is negligible in terms of power and area.

TABLE 4
OVERHEAD OF ASYMMETRIC AGING-AWARE SCHEME FOR EXECUTION UNITS

| Area [um²] | Power [uW] | Timing impact [ps] |
|---|---|---|
| 872 | 107 | 80 |

## B. Architectural Registers

The experimental observations presented in TABLE 2 clearly indicate that there is a significant number of registers that incur BTI stress. Typically, out-of-order microprocessor architectural registers are renamed as physical ones and are hosted by a cyclic buffer (as part of the RoB). This implementation mitigates BTI stress for the physical registers since architectural register mapping changes their physical locations rapidly. Nonetheless, architectural registers, and in particular static control and configuration registers, such as CRs and MSRs, still suffer from asymmetric aging. The proposed architectural solution, illustrated in Fig. *5*, avoids BTI hotspots by periodically changing the mapping of registers to their corresponding architectural hosting locations.

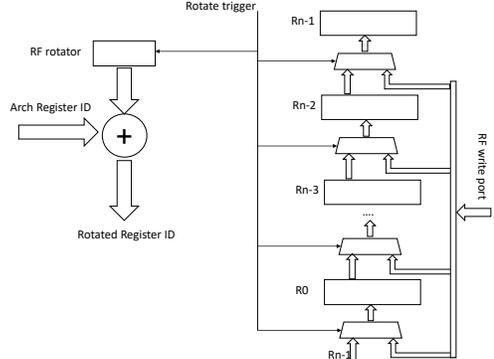

Fig. 5 Asymmetric aging-aware register mapping scheme.

The scheme is based on modulo rotation of the mapping between the architectural or control register identifiers and their physical locations. As illustrated in Fig. *3*, a pulse trigger is asserted to shift the register mapping in the register file (RF) either periodically at low frequency (or each time we change CR3) or as part of the return-from-interrupt procedure before saving the values of the user-level process. A modulo-counter (RF rotator) serves to map the architectural or control register number to the mapped register location by modulo addition. After each assertion of the rotation trigger (at any arbitrary time point), the counter is incremented, and the register values are shifted between registers, as illustrated in Fig. *5*. When examining the proposed scheme in our simulation environment, we observe that it is able to prevent the static stress that was reported in TABLE 2. The rotation trigger in our simulation was asserted every 10 million clock cycles. We examined different rotation trigger rates and found that this value does not impact performance. Table 5 summarizes the power, timing path and gate count overhead for 28nm-process technology (for a bulk of 32 registers) using HVT standard cells. As can be observed, the overall overhead is very small:

TABLE 5
ASYMMETRIC AGING-AWARE REGISTERS MAPPING SCHEME OVERHEAD

| Area [um²] | Power [uW] | Timing impact [ps] |
|---|---|---|
| 1973 | 190 | 80 |

The proposed solution may appear to resemble the Sun SPARC and Berkley RISC CPU register window [47], which is used for a different purpose. Register windows are a scheme that aims to evenly distribute sets of GPR registers between different sections of code, typically procedure calls. At every nested call, the register window is shifted to provide the program with a new working set of registers. In contrast, our proposed scheme extends to all architectural registers (FP, vector, control etc.) and shifts one register at a time unlike, the register window technique that shifts a bulk of registers and is limited to integer registers. It should also be noted that register windows involve more frequent register window switches, resulting in excessive dynamic power while the rotation frequency of our proposed scheme is very low.

### C. Memory Hierarchy

Our observations, presented in Fig. 2, suggest that cache entries may incur BTI stress in different levels of the memory hierarchy. Again, one may argue that in general purpose, multiprocessing environments such phenomena may be rare due to the frequent context switching. Our experiments, however, indicate that this situation is quite common for system environments that run the same program for very long periods (e.g., embedded microcontrollers) since they cannot leverage any context switching to purge constant values out of the cache memories. Our proposed technique can mitigate major reliability concerns regarding such systems. We suggest combining a PRBS generator with the swap-shift set index remapping method, introduced by Wang et al. ([48]). The swap-shift method was proposed to handle a different reliability phenomenon, termed write endurance, in PCM-based non-volatile memories (NVMs). Unlike NBTI where the stress is induced by constant values maintained for long periods, PCM-based NVMs experience bit cell wear-out after an excessive number of writes, leaving the bit cell resistance in a low or high resistance state.

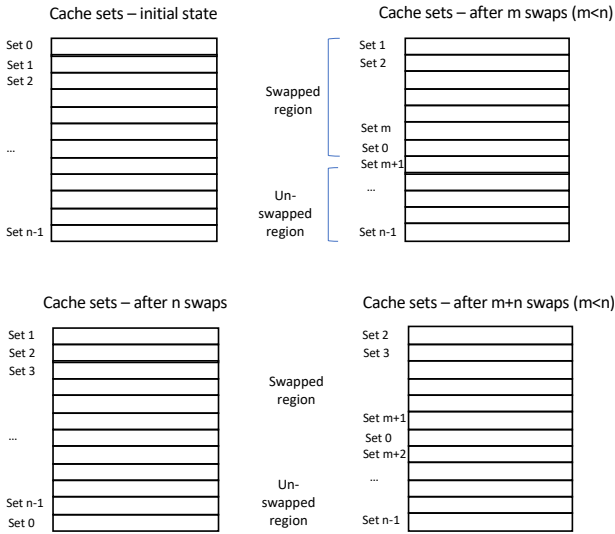

Fig. 6 The swap-shift scheme for cache memories.

The original swap-shift technique, illustrated in Fig. 6, suggests periodically swapping a pair of cache sets whenever a number of writes reaches a certain threshold. It should be noted that the swapping only changes the set identifier, and it invalidates the data residing in the swapped pair of lines. A swapped-set counter is maintained through the shift process to identify the last set that was shifted. This counter is incremented whenever a set is swapped. Once all sets are swapped, the swapped-set counter wraps to 0, the cache index mapping is rotated by one and the whole process is restarted. The accumulative number of cache index rotations is maintained by the set-shift counter and once it reaches the number of sets it wraps to 0.

We propose employing a modified swap-shift scheme combined with a PRBS generator as illustrated in Fig. 7. The swap-shift process is activated by a periodic shift trigger that is asserted in low frequency. The set-shift and the set-swap counters operate in similar fashion to the original swap-shift method except for the fact that the set swapping is triggered by the periodic shift trigger signal. Upon any swapping, a pseudorandom pattern, generated by the PRBS module, is written into the invalidated swapped cache sets. To minimize the timing impact on the cache access time, we retime the index remapping logic and place it in a pipeline stage prior to the cache access stage (address generation and memory order buffer). The usage of the combined swap-shift and PRBS generator serves to eliminate constant BTI stress induced by two scenarios: 1. Constant values residing in the cache for long periods – In this case the set rotation spreads the BTI stress uniformly across all sets. 2. Unutilized cache lines that remain unchanged for long periods despite the set shift – In this scenario, the PRBS generator helps eliminate the BTI stress even if the physical set remains unutilized after the swapping.

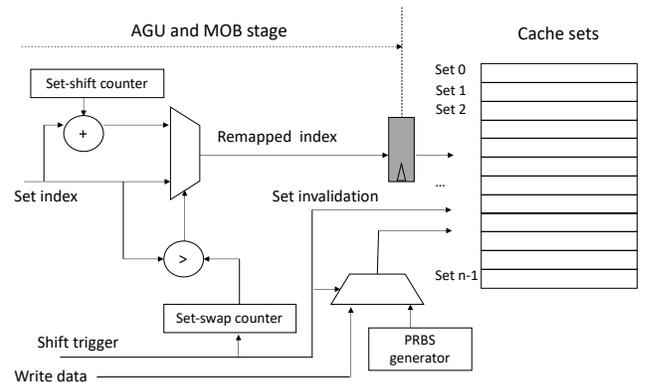

Fig. 7 Asymmetric aware cache memory sets remapping.

The proposed scheme introduces several advantages in respect to CFDR and the scrambling-probing methods. First, it eliminates the flushing of the entire cache and hence minimizes the performance overhead. The proposed technique was examined in our experimental environment with various shift-trigger rates and it was found that a shift-trigger every 10M-cache accesses yields minimal performance overhead (less than 0.01% impact on cycle count) while fully eliminating BTI stress of constant values. In addition, our scheme introduces simpler logic complexity due to the avoidance of read-modify-write sequences required by the CFDR method. Such a sequence overloads the cache port with a significant timing overhead and may require sampling stages and an additional dedicated FSM to maintain such a flow. Last, we minimize the timing impact on the cache access time by retiming the index remapping to prior pipeline stages. Our logical scheme was synthesized, and Table 6 summarizes power, timing and area overhead obtained by the synthesis for 28nm-process technology. Since the propagation delay of the index remapping in our proposed scheme may be significant, we present our synthesis results using different threshold voltage standard cell options: HVT, standard VT (SVT), low VT (LVT) and ultra-low VT (uLVT). This allows us to find the optimal tradeoff for the propagation delay given system power constraints. It can be observed that the area overhead in very small. In addition, the index propagation delay can be reduced by more than x3 when using uLVT in respect to HVT cells. The power, however, increases by approximately x50–60. SVT and LVT cells introduced a nearly x1.4 and x2, respectively, propagation delay improvement in respect to HVT cells but with a power overhead significantly smaller than in the uLVT cells.

TABLE 6
ASYMMETRIC AGING-AWARE CACHE MEMORY OVERHEAD

| Cache | Standard cells VT | Area [mm²] | Power [uW] | Delay to block data [ns] | Delay to index [ns] |
|---|---|---|---|---|---|
| L1-D | HVT | | 865 | 0.08 | 0.35 |
| | SVT | | 6170 | 0.06 | 0.25 |
| | LVT | 0.015 | 19050 | 0.04 | 0.18 |
| | uLVT | | 48600 | 0.02 | 0.10 |
| L1-I | HVT | | 505 | 0.08 | 0.35 |
| | SVT | | 3290 | 0.06 | 0.25 |
| | LVT | 0.008 | 10050 | 0.04 | 0.18 |
| | uLVT | | 25600 | 0.02 | 0.10 |
| L2 | HVT | | 867 | 0.08 | 0.40 |
| | SVT | | 6283 | 0.06 | 0.29 |
| | LVT | 0.015 | 19400 | 0.04 | 0.21 |
| | uLVT | | 49500 | 0.02 | 0.12 |
| L3 | HVT | | 1627 | 0.08 | 0.40 |
| | SVT | | 12270 | 0.06 | 0.29 |
| | LVT | 0.028 | 38110 | 0.04 | 0.21 |
| | uLVT | | 97400 | 0.02 | 0.12 |

## V. CONCLUSIONS AND DISCUSSION

Microprocess reliability is a crucial requirement that has been highly challenged by advanced process technologies and new computation-intensive applications such as autonomous vehicles, data centers, cloud computing and life-support systems. Recent advanced process nodes have become highly susceptible to asymmetric aging that can cause critical timing violations in ICs and overall system failure. Asymmetric aging is primarily induced by the BTI effect when constant voltage is applied to transistor gates for long duration. We summarize the contributions offered in this paper as follow:

1) We examined the asymmetric aging root cause in microprocessors and identified three new main sources of BTI stress. First, backward compatible features may lead to underutilization and BTI stress. Second, new forward compatible features—when not efficiently utilized due to legacy consideration—may also prompt asymmetric aging concerns. Third, microprocessors are by nature general purpose machines and, as a result, it is highly challenging to inhibit static BTI stress on all their logical circuits.

2) We introduced a novel mechanism to avoid asymmetric aging in execution units that are under BTI stress through periodic injection of pseudorandom data at low rates.

3) We presented a mechanism for asymmetric aging avoidance in architectural and control registers by employing periodical shift and remapping of register identifiers.

4) We offered a mechanism to handle asymmetric aging in memory systems using a new approach that combines swap-shift of cache sets and a pseudorandom data generator.

5) Our experimental results indicated that the proposed techniques were able to efficiently eliminate constant BTI stress with negligible performance impact. In addition, we performed a synthesis trial and examined power, timing and area impact of the proposed mechanism and found that all methods introduce very small overhead.

Asymmetric transistor aging is becoming a highly important phenomenon in many fields such as embedded systems, autonomous cares, memory data bases, and more. Many of these environments require system architects to guarantee the lifetime of products, which may be governed by their reliability. Meeting such demand requires further extensive studies by different disciplines: process technology, physical design, EDA tools and system microarchitecture. In the most advanced process technologies of 5 and 3nm, reliability related issues are expected to become even more complex mainly because the HCI effect becomes more dominant there. Physical design flows should be developed to better analyze and fix asymmetric aging violations in large-scale circuits. This is a major challenge that requires both industry and research communities to find practical solutions to allow the development of future reliable large-scale ICs. System architects are also encouraged to conduct asymmetric aging studies in order to mitigate this phenomenon by architectural means. Additional processing systems such as GPUs, FPGAs, networking systems and dedicated processing accelerators should be further examined to find innovative architectural solutions for asymmetric aging mitigation.

## REFERENCES

[1] Operating Temperature, Wikipedia https://en.wikipedia.org/wiki/Operating_temperature.

[2] Failure Mechanism Based Stress Test Qualification for Integrated Circuit. Automotive Electronics Council, Component Technical Committee – AEC – Q100 – REV-G standard.

[3] M. A. Alam, H. Kufluoglu, D. Varghese, and S. Mahapatra, "A comprehensive model for PMOS NBTI degradation," *Microelectron. Rel.*, vol. 47, no. 6, pp. 853–862, Jun. 2007. https://doi.org/10.1016/j.microrel.2006.10.012

[4] S. Bharadwaj, W. Wang, R. Vattikonda, Y. Cao, and S. Vrudhula, "Predictive modeling of the NBTI effect for reliable design," in *Proc. Custom Integrated Circuits Conf.*, Sep. 2006, pp. 189–192.

[5] W. Wang, V. Reddy, A. T. Krishnan, R. Vattikonda, S. Krishnan, and Y. Cao, "Compact modeling and simulation of circuit reliability for 65 nm CMOS technology," IEEE Trans. Device Mater. Rel., vol. 7, no. 4, pp. 509–517, Dec. 2007.

[6] A. Calimera, M. Loghi, E. Macii, and M. Poncino, "Aging effects of leakage optimizations for caches," in Proc. IEEE Great Lakes Symp. VLSI, May 2010, pp. 95–98.

[7] J. B. Velamala, K. B. Sutaria, V. S. Ravi, and Y. Cao. Failure analysis of asymmetric aging under NBTI," IEEE Trans. Device Mater. Rel., vol. 13, no. 2, pp. 340–349, June 2013.

[8] A. Campos-Cruz, G. Espinosa-Flores-Verdad, A. Torres-Jacome, and E. Tlelo-Cuautle, On the prediction of the threshold voltage degradation in CMOS technology due to bias-temperature instability. Electronics, vol. 7, no. 12, pp. 427 2018, 7, 427.

[9] M. Taddiken, N. Hellwege, Nils Heidmann, Dagmar Peters-Drolshagen and Steffen Paul. Analysis of aging effects - From transistor to system level. Microelectronics Reliability Volume 67, December 2016, Pages 64-73

[10] Hussam Amrouch, Behnam Khaleghi, Andreas Gerstlauer and Jörg Henkel. Reliability-Aware Design to Suppress Aging. DAC '16: Proceedings of the 53rd Annual Design Automation Conference. June 2016. Article No.: 12 Pages 1–6 https://doi.org/10.1145/2897937.2898082

[11] Omer Khan and Sandip Kundu. A Self-Adaptive System Architecture to Address Transistor Aging. 2009 Design, Automation & Test in Europe Conference & Exhibition.

[12] A. T. Krishnan et al., "Product drift from NBTI: Guardbanding, circuit and statistical effects," 2010 International Electron Devices Meeting, San Francisco, CA, 2010, pp. 4.3.1-4.3.4.

[13] R. Zheng, J. Velamala, V. Reddy, V. Balakrishnan, E. Mintarno, S. Mitra, S. Krishnan, and Y. Cao, "Circuit aging prediction for low-power operation," in Proc. Custom Integr. Circuits Conf., Sep. 2009, pp. 427–430.

[14] HTOL - Temperature, Bias, and Operating Life. JEDEC JESD22-A108F standard.

[15] S. Ogawa and N. Shiono, "Generalized diffusion-reaction model for the low-field charge build up instability at the Si-SiO$_2$ interface," Physical Review, 51(7):4218–4230, Feb. 1995.

[16] K. L. Chen, S. A. Saller, I. A. Groves, C. C. Li, E. Minami, and D. B. Scott, "Reliability effects on MOS transistors due to hot-carrier injection," IEEE Trans. Electron Devices, vol. 32, no. 2, pp. 386–393, Feb. 1985.

[17] Reliability Simulation in Integrated Circuit Design. [Online]. Available: http://www.cadence.com

[18] Jyothi Bhaskarr Velamala. Compact Modeling and Simulation for Digital Circuit Aging. PhD dissertation. https://repository.asu.edu/attachments/97628/content/tmp/package-sAOTIT/Velamala_asu_0010E_12271.pdf

[19] R. Vattikonda, W. Wang, and Y. Cao, "Modeling and minimization of pmos nbti effect for robust nanometer design", IEEE/ACM Design Automation Conference, pages 1047–1052, Jul. 2006.

[20] J. KO and S. CM, "Negative bias stress of mos devices at high electric fields and degradation of mos devices", Journal of Applied Physics, 48(5):2004– 2014, May 1977.

[21] M. Agarwal, B. C. Paul, Ming Zhang, and S. Mitra, "Circuit failure prediction and its application to transistor aging", VLSI Test Symposium, pages 277–286, May 2007.

[22] W. Wang, Z. Wei, S. Yang, and Y. Cao, "An efficient method to identify critical gates under circuit aging," in Proc. Int. Conf. Comput. Aided Des., Nov. 2007, pp. 735–740.

[23] B. C. Paul, K. Kang, H. Kufluoglu, M. A. Alam, and K. Roy, "Impact of NBTI on the temporal performance degradation of digital circuits," IEEE Electron Device Lett., vol. 26, no. 8, pp. 560–562, Aug. 2005.

[24] W. Wang, S. Yang, S. Bhardwaj, R. Vattikonda, S. Vrudhula, F. Liu, and Y. Cao, "The impact of NBTI effect on combinational circuit: Modeling, simulation, and analysis," IEEE Trans. VLSI Syst., vol. 18, no. 2, pp. 173– 183, Feb. 2010.

[25] H. Sangwoo and K. John, "NBTI-aware statistical timing analysis framework," in Proc. IEEE Int. SOC Conf., 2010, pp. 158–163.

[26] A. Ricketts, J. Singh., K. Ramakrishnan, N. Vijaykrishnan, and D. K. Pradhan, "Investigating the impact of NBTI on different power saving cache strategies," in Proc. DATE, Mar. 2010, pp. 592–597.

[27] M. Powell, S.-H. Yang, B. Falsafi, K. Roy, and T. N. Vijaykumar, "Gated-VDD: A circuit technique to reduce leakage in deep-submicron cache memories," in Proc. ISLPED, Jul. 2000, pp. 90–95.

[28] S. Kaxiras, Z. Hu, and M. Martonosi, "Cache decay: Exploiting general behavior to reduce cache leakage power," in Proc. ISCA, Jun. 2001, pp. 240–251.

[29] K. Flautner, N. Kim, S. Martin, D. Blaauw, and T. Mudge, "Drowsy caches: Simple techniques for reducing leakage power," in Proc. ISCA, May 2002, pp. 148–157.

[30] X. Wang, W. Xu and C. H. Kim, "SRAM read performance degradation under asymmetric NBTI and PBTI stress: Characterization vehicle and statistical aging data," Proceedings of the IEEE 2014 Custom Integrated Circuits Conference, San Jose, CA, 2014, pp. 1-4.

[31] J. L Yan, X. J. Li and Y. L. Shi. The impact of Negative Bias Temperature Instability (NBTI) effect on D flip-flop. Proceedings of the 2014 Asia-Pacific Conference of Electronics and Electrical Engineering.

[32] F. Firouzi, S. Kiamehr and M. B. Tahoori, "NBTI mitigation by optimized NOP assignment and insertion," 2012 Design, Automation & Test in Europe Conference & Exhibition (DATE), Dresden, 2012, pp. 218-223.

[33] Haider Muhi Abbas, Mark Zwolinski, and Basel Halak. Aging Mitigation Techniques for Microprocessors Using Anti-aging Software. Chapter 3, Ageing of Integrated Circuits - Causes, Effects and Mitigation Techniques, Springer, Cham. ISBN 978-3-030-23781-3

[34] Y. Chen, I. Lin and J. Ke, "ROAD: Improving Reliability of Multi-core System via Asymmetric Aging," 2019 IEEE/ACM International Conference on Computer-Aided Design (ICCAD), Westminster, CO, USA, 2019, pp. 1-8.

[35] Techniques for Reducing Uneven Aging in Integrated Circuits. Intel European Patent Application. Application number 18190097.8. 2018.

[36] M. S. Mispan, M. Zwolinski, and B. Halak. Ageing Mitigation Techniques for SRAM Memories. Chapter 4, Ageing of Integrated Circuits - Causes, Effects and Mitigation Techniques, Springer, Cham. ISBN 978-3-030-23781-3

[37] Kumar, S. V., Kim, C. H., & Sapatnekar, S. S. (2006). Impact of NBTI on SRAM read stability and design for reliability. In: International symposium on quality electronic design (pp. 210–218).

[38] Gebregiorgis, A., Ebrahimi, M., Kiamehr, S., Oboril, F., Hamdioui, S., & Tahoori, M. B. (2015). Aging mitigation in memory arrays using self-controlled bit-flipping technique. In: Asia and South Pacific design automation conference (pp. 231–236).

[39] Duan, S., Halak, B., & Zwolinski, M. (2018). Cell flipping with distributed refresh for cache ageing minimization. In: IEEE Asian test symposium (pp. 1–6).

[40] A. Calimera, M. Loghi, E. Macii and M. Poncino, "Dynamic Indexing: Leakage-Aging Co-Optimization for Caches," in IEEE Transactions on Computer-Aided Design of Integrated Circuits and Systems, vol. 33, no. 2, pp. 251-264, Feb. 2014, doi: 10.1109/TCAD.2013.2287187

[41] T. E. Carlson, W. Heirman, and L. Eeckhout. Sniper: Exploring the level of abstraction for scalable and accurate parallel multi-core simulations. In Proceedings of the International Conference for High Performance Computing, Net- working, Storage and Analysis (SC), Nov. 2011.

[42] M. E. Thomadakis. The architecture of the Nehalem processor and Nehalem-EP smp platforms. Technical report, December 2010. http://sc.tamu.edu/systems/eos/nehalem.pdf.

[43] A. Limaye and T. Adegbija, "A workload characterization of the spec cpu2017 benchmark suite," in 2018 IEEE International Symposium on Performance Analysis of Systems and Software (ISPASS), pp. 149–158, April 2018

[44] Q. Wu, Steven Flolid, Shuang Song, Junyong Deng, Lizy K. John. Hot Regions in SPEC CPU2017. 2018 IEEE International Symposium on Workload Characterization (IISWC).

[45] H. Esmaeilzadeh, E. Blem, R. St. Amant, K. Sankaralingam, and D. Burger. 2011. Dark silicon and the end of multicore scaling. In Proceedings of the 38th annual international symposium on Computer architecture (ISCA '11). Association for Computing Machinery, New York, NY, USA, 365–376. DOI:https://doi.org/10.1145/2000064.2000108

[46] L. Pavlovic, M. Vidmar and S. Tomazic, "2.5 Gbit/s PRBS Generator and Checker," Proceedings ELMAR 2006, Zadar, 2006, pp. 363-366, doi: 10.1109/ELMAR.2006.329585.

[47] The SPARC Architecture Manual, Version 8.

[48] J. Wang, X. Dong, Y. Xie and N. P. Jouppi, "i2WAP: Improving non-volatile cache lifetime by reducing inter- and intra-set write variations," 2013 IEEE 19th International Symposium on High Performance Computer Architecture (HPCA), Shenzhen, 2013, pp. 234-245, doi: 10.1109/HPCA.2013.6522322.

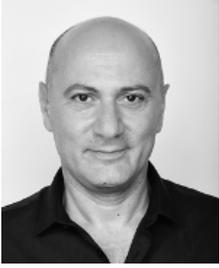

**Freddy Gabbay** received his B.Sc., M.Sc. and Ph.D. in electrical engineering from the Technion – Israel Institute of Technology, Haifa, Israel in 1994, 1995 and 1998, respectively. His areas of research are HPC accelerators, VLSI design, chip reliability, microprocessor architecture and machine learning.

In 1998, he worked as a researcher at Intel's Microprocessor Research Lab. In 1999 he joined Mellanox Technologies and held various positions in leading diverse product line architectures and ASIC design. In 2003, he joined Freescale Semiconductor as a senior design manager and led the design of baseband ASIC products. In 2012 he rejoined Mellanox Technologies where he served as Vice President of Chip Design.

Today he is an Associate Professor at Ruppin Academic Center, Israel. Prof. Gabbay also holds 19 patents.

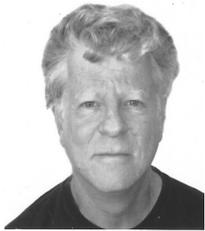

**Avi Mendelson** is a visiting professor at the CS and EE Departments of the Technion – Israel Institute of Technology, Haifa, Israel and in the EE Department of NTU, Singapore. He has a blend of industrial and academic experience in several different areas such as computer architecture, power management, security, and real-time systems.

As part of his industrial role, he worked for National Semiconductor, on the team that invented and developed the first PC-on-Chip. At Intel he worked for five years as a researcher in Intel Research Labs and six years as principle engineer in the mobile CPU architecture team where he was chief architecture of the first CMP feature (multicore) of Intel cores. For this task and leadership, he received the IAA (Intel Achievement Award).

Prof. Avi Mendelson is an IEEE Fellow, was a member of the Board of Governors of the IEEE Computer Society and served as a second VP of the IEEE Computer Society.